# Metamaterials with interacting Metaatoms

A. Chipouline, S. Sugavanam, J. Petschulat, T. Pertsch

Institute of Applied Physics, Friedrich-Schiller-Universität Jena, Max Wien Platz 1, 07743, Jena, Germany

**Abstract**

An analytical model for homogenization of Metamaterials (MM) with interacting Metaatoms (MA) is developed based on multipole approach for bulk media. The interaction is assumed to be near field type. i.e. no retardation in lateral direction between adjacent MAs is assumed. The interaction takes place between the adjacent MAs only in lateral (perpendicular to the propagation) direction; the considered MM is supposed to consist of non-interacting identical layers with periodically spaced MAs. It is shown, that the interaction between MAs leads to significant changes of dispersion characteristics, in particular, appearance of the spatial dispersion is emphasized. The results indicate increase of the available real part of the *k* vector values and increase of imaginary part by the approximately same percentage, which leads to the final conclusion that the effect of coupling does not provide extra opportunities for the resolution enhancement.

1. **Introduction**

Volume or statistical averaging of the microscopic Maxwell equations (MEs), i.e. transition from microscopic MEs to their macroscopic counterparts, is one of the main steps in electrodynamics of materials. In spite of the fundamental importance of the averaging procedure, it is quite rarely properly discussed in university courses and respective books; up to now there is no established consensus about how the averaging procedure has to be performed. In [1] it has been show that there are some basic principles for the averaging procedure (irrespective to what type of material is studied) which have to be satisfied in order to be accepted as a credible one. In this paper show how the multipole approach in homogenization of the MM [2, 3] can be extended on the case of regularly placed interacting MA in form of the double wire structure [4].

The multipole expansion and effective medium constants elaboration, developed in [5], result in constructive expressions for multipoles as functions of microscopic charge dynamics. This solves the problem of homogenization, provided the microscopic charge dynamics is expressed through the averaged (macroscopic) electric and magnetic fields. Nevertheless, during the elaboration of this approach, a lot of assumptions have to be accepted, and many of them can be rather poor justified for the typical MM configurations. For example, assumptions about significantly large distance between atoms in compare with the atomic sizes is undoubtedly good satisfied in case of solid states, but for MM, where MA are just maximum five times smaller (for typical structures) then the distance between them, this approximation is rather questionable.

Nevertheless, it is believed that the approach used in [3] is satisfactory good (at least as a first approximation); moreover, the fitting parameters used in the model absorbs to some extend the mentioned above deviation between rigorous representation [5] and one elaborated in [3].

The approach presented in [3] has a lot of potential for further development due to the fact that the expressions for microscopic dynamics of charges can be elaborated, taking into account interaction between MA (which is a subject of this paper), and/or MA of more complex structure, consisting of not only plasmonic nano resonators, but, for example, the plasmonic nano resonators coupled with quantum systems, like Carbon Nanotubes [6] or Quantum Dots [7].

In this paper we consider a metamaterial consisting of the identical layers with regularly spaced MA in each; one layer, which the MM consists of (the layers repeat themselves in *y* direction), is presented in Fig. 1.

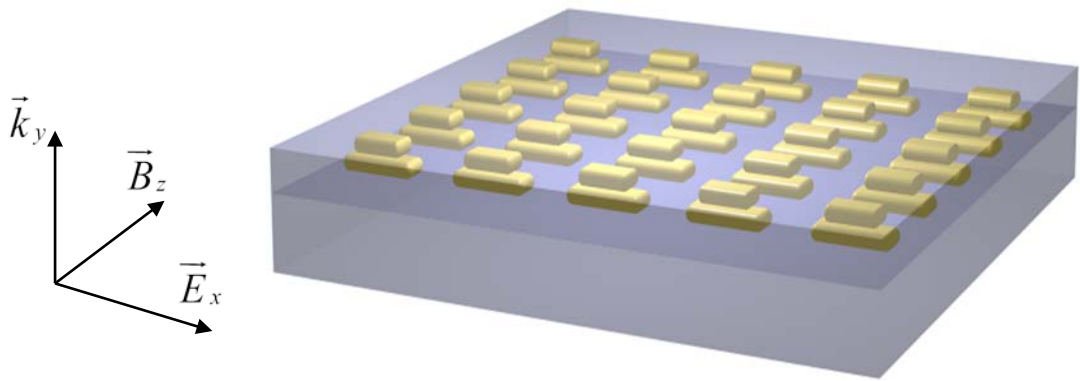

**Fig. 1: Artificial metaatoms (plasmonic nano resonators) embedded in a dielectric matrix form a MM (only one layer is presented). Polarization of the electric and magnetic fields, and direction of the wave vector are shown. The interaction between MAs takes place in *z* direction; the possible interaction in *y* direction (wave propagation direction) is not taken into account.**

The interaction between MAs is assumed to be negligible in longitudinal direction (perpendicular to the layer surfaces) or, in other words, the layers are assumed to be good separated from each other. The effect of interaction in lateral direction (parallel to the layer surface) and its influence on the dispersion relation of the plane waves propagating in the MM consisting of the interacting MAs, is a subject of the presented paper.

When there is a coupling between MA, the response of the medium is no longer truly local. As a result, depending upon the configuration of the MA, the medium responds differently to electromagnetic waves propagating in different directions. This phenomenon is called spatial dispersion (akin to the phenomenon of temporal dispersion – where the response of a medium at a given time depends on the history of its excitation). The direction dependence of the medium response is not just unique for the spatial dispersion – in case of anisotropic media this effect appears as well. The qualitative markers for differentiation of these two effects has been considered in details in [9] and [10] (see also references therein). First, the effect of anisotropy which leads to different wave vectors for different propagation directions and which is typical for crystals, has to be taken into account. It was shown that the effect of anisotropy leads to two possible families of isofrequency contours, namely ellipsoids and hyperboloids, depending on the sign of the ratios $\frac{\varepsilon_{zz}}{\varepsilon_{xx}}$ and $\frac{\varepsilon_{yy}}{\varepsilon_{xx}}$ (here the consideration is carried out in the main axis of the intrinsic coordinate system of the crystal) [11], [12], [13]; the effective parameters $\varepsilon_{xx}$, $\varepsilon_{yy}$, and $\varepsilon_{zz}$ themself do not depend on the wave vector (spatial dispersion is negligible). In media with spatial dispersion the effective parameters depend on the wave vector considerably, therefore making the shape of the isofrequency contours arbitrary. Comparison of isofrequencies of MM obtained numerically with ellipsoids and hyperboloids makes it possible to identify frequency intervals, where the effective parameters do not depend on the wave vector.

One of the ways to describe spatial dispersion is to use a model of a chain of the coupled harmonic oscillators. Such a model is adequate as a first approximation for the interactions between plasmonic oscillations in the metallic nano-resonators. Eigen modes of the response are obtained as wave solutions giving the oscillations of the plasmonic charges (see Fig. 2).

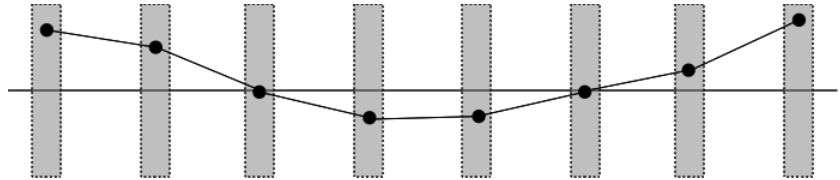

**Fig. 2: Spatial dispersion viewed as a consequence of a coupling effect in a chain of dipoles. The problem is equivalent to the study of transverse oscillation dynamics in a chain of the coupled harmonic oscillators.**

Similar approach has been used to study effect of interacting of the MAs on the properties of the MM in [14-17] for microwave frequency region.

To better understand the problem, first the case of a one dimensional chain of coupled harmonic oscillators (coupled dipoles) is studied, after which the problem is extended to the case of coupled MA in form of cut-wires (coupled quadrupoles). The results of the analysis for both ensembles are presented side by side to enable a comparative study of the dispersion characteristics.

The knowledge of the dispersion relation is very important; actually, it is the only, what is required to analyze the wave propagation in a media (boundary condition problems are not included in the discussion here). For example, it is known that in order to provide a better resolution the media has to allow propagation of the lateral components of the wave vector $\vec{k}$ with as much as possible values (as higher as possible $k_z$ components in Fig. 3 for the same wavelength), which could be achieved, for example, in a hyperbolic dispersion media [18]. The performed below analysis provides a tool to analyze whether the coupling between MAs could enhance the available spatial spectrum of the propagating waves and hence increase resolution of the optical systems, which use the respective MM.

The paper consists of five parts. After introduction (part 1), the dispersion relations for eigen waves in chains of coupled dipoles and quadrupoles will be summarized in part 2. The dispersion relations in form of transcendental equations for the electro-magnetic plane waves propagating in the media, consisting of the coupled dipoles and quadrupoles will be elaborated in part 3. In part 4 the respective solutions will be provided and the results will be summarized; short conclusion is given in part 5.

2. **Dispersion relations for material eigen waves**
2.1 **Periodic chain of coupled dipoles**

A chain of periodically positioned dipoles (oriented with the long axis along the *x* direction) is considered (see Fig. 3). For clarity, only one row is shown in the figure; it is assumed that rows of dipoles are placed along the *y* direction. The treatment of the problem remains the same, and coupling between neighboring rows are neglected. The arrangement of the dipoles is along the *z* direction; the period of spacing is taken to be $z_0$. The effect of coupling with adjacent oscillators is introduced via a coupling constant $\sigma$ that is a function of the distance of separation between the oscillators. Well known solution in form of the transverse spatial modes which can be sustained in such a medium under the above mentioned conditions can be obtained straightforwardly.

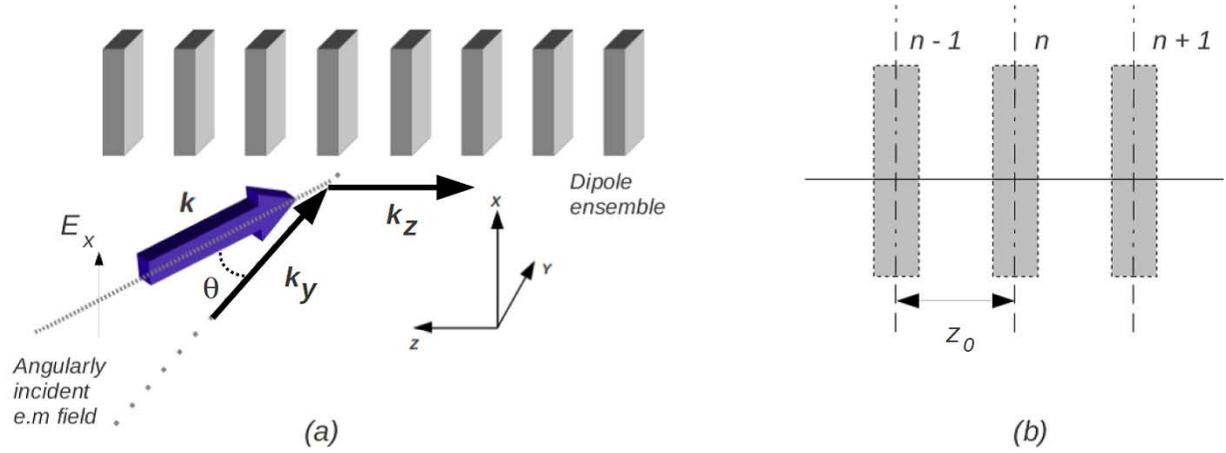

**Fig. 3: Geometry of propagation** - (a) The electric field is polarized along the long axis of the cut-wires; angular incidence gives rise to spatial modes in the ensemble. (b) The dipoles in an arbitrary triplet are labeled as n, n+1 and n-1. The positional coordinates of the charge clouds $x_n$ within these dipoles are also indexed with these labels.

The dynamic equation for the $n-th$ oscillator is:

$$\frac{\partial^2 x_n}{\partial t^2} + \gamma \frac{\partial x_n}{\partial t} + \omega_0^2 x_n + \sigma(x_{n+1} + x_{n-1}) = \frac{q}{m} E_x \qquad (1)$$

The allowed modes can then be obtained by transferring the problem to the Fourier domain using the ansatz:

$$x_n = A_0 \exp(ik_z n z_0 - i\omega t) \qquad (2)$$

The amplitude of the $n-th$ oscillator in terms of the wave vector $k_z$ is thus given by:

$$x_n = A_0 \exp(ik_z n z_0) \qquad (3)$$

Essentially then, the ansatz describes a system of oscillators vibrating at the same frequency, with *k* giving the periodicity of the spatial mode. Substitution of this ansatz in (1) gives the dispersion relation:

$$\omega_0^2 - \omega^2 - i\omega\gamma + 2\sigma\cos(k_z z_0) = 0 \qquad (4)$$

The solution of this equation is presented in Figs. 8 and 9 and the respective discussion is given in part 4.

### 2.2 Periodic chain of coupled quadrupoles

The above arguments are now extended to the case of the chain of coupled quadrupoles. The two cut-wires forming the quadrupole are assumed to be oriented with their long axes along the *x* direction as before, while being spaced by a small distance $2y_1$ apart along the y direction. It is further assumed that the quadrupoles themselves are periodically spaced along the *z* direction, with the spatial period $z_0$. The interaction of a single cut-wire with five of its nearest neighbors has to be considered (see Fig. 4) for both cut-wires (that is, both $x_n$ and $x_n'$) as they experience different excitation conditions due to the retardation of the wave propagating into *y* direction. As before, we are looking for the transverse spatial modes which can be sustained in such a medium. The dynamic equations for the two oscillators can be written as:

$$\begin{cases} \dfrac{\partial^2 x_n}{\partial t^2} + \gamma \dfrac{\partial x_n}{\partial t} + \omega_0^2 x_n + \sigma\left(x_{n+1} + x_{n-1}\right) + \sigma'\left(x'_{n+1} + x'_{n-1}\right) + \sigma_0 x'_n = 0 \\ \dfrac{\partial^2 x'_n}{\partial t^2} + \gamma \dfrac{\partial x'_n}{\partial t} + \omega_0^2 x'_n + \sigma\left(x'_{n+1} + x'_{n-1}\right) + \sigma'\left(x_{n+1} + x_{n-1}\right) + \sigma_0 x_n = 0 \end{cases} \quad (5)$$

where $x_n$ and $x'_n$ are the instantaneous coordinates of the plasmonic charge clouds on the cut-wires, while $\sigma_0$, $\sigma$, and $\sigma'$ are the coupling coefficients:

$$\sigma = \sigma_0 \frac{(2y_1)^3}{z_0^3}, \quad \sigma' = \sigma_0 \frac{(2y_1)^3}{\left(z_0^2 + (2y_1)^2\right)^{3/2}} \quad (6)$$

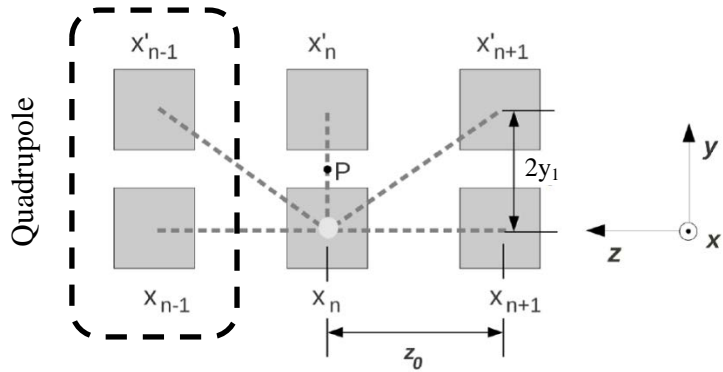

**Fig. 4: Nearest neighbor interactions - top view of a one dimensional chain of the quadrupoles (two cut-wires forming one from the three shown quadrupoles are surrounded by dashed frame). The dashed lines indicate the interactions that have to be taken into account.**

Assuming the ansatz:

$$x_n = A_0 \exp(ik_z n z_0 - i\omega t), \quad x'_n = A'_0 \exp(ik_z n z_0 - i\omega t) \quad (7)$$

where $2y_1$ is the spatial period of the quadrupoles in the direction of the wave propagation $y$. Substituting the ansatz in the dynamic equations, we arrive at:

$$\begin{cases} A_0\left(R + 2\sigma \cos(k_z z_0)\right) + A'_0\left(2\sigma' \cos(k_z z_0) + \sigma_0\right) = 0 \\ A_0\left(2\sigma' \cos(k_z z_0) + \sigma_0\right) + A'_0\left(R + 2\sigma \cos(k_z z_0)\right) = 0 \end{cases} \quad (8)$$

here $R = \omega_0^2 - \omega^2 - i\gamma\omega$. For the above system of equations to be consistent, the determinant must vanish:

$$\begin{vmatrix} R+2\sigma\cos(k_z z_0) & 2\sigma'\cos(k_z z_0)+\sigma_0 \\ 2\sigma'\cos(k_z z_0)+\sigma_0 & R+2\sigma\cos(k_z z_0) \end{vmatrix}=0 \qquad (9)$$

This gives the dispersion relation equation:

$$\left(R+\sigma_0+2(\sigma+\sigma')\cos(k_z z_0)\right)\left(R-\sigma_0+2(\sigma-\sigma')\cos(k_z z_0)\right)=0 \qquad (10)$$

which results in two equations, which give the bands of frequencies capable of producing the spatial modes. The equations are:

$$\begin{cases} \omega_0^2-\omega^2-i\omega\gamma+\sigma_0+2(\sigma+\sigma')\cos(k_z z_0)=0 \\ \omega_0^2-\omega^2-i\omega\gamma-\sigma_0+2(\sigma-\sigma')\cos(k_z z_0)=0 \end{cases} \qquad (11)$$

The solutions of the above equations are presented in Figs. 8 and 9 and the respective discussion is given in part 4.

### 3. Dispersion relations for electromagnetic waves

### 3.1 Periodic chain of coupled dipoles

To obtain the dispersion relation for the electromagnetic wave propagating in a media with periodic chains of coupled dipoles, the Helmholtz equation has to be employed. The electric field is assumed to be in the form:

$$E_x=E_{x,0}\exp(ik_y y+ik_z n z_0-i\omega t) \qquad (12)$$

which is a plane wave polarized along the $x$ axis, with its wave vector in the $yz$ plane. The propagation vectors along $y$ and $z$ directions are denoted by $k_y$ and $k_z$. In turn, $k_y$ and $k_x$ can be expressed in their

polar forms:

$$k_y = k\cos(\theta)$$
$$k_z = k\sin(\theta) \quad (13)$$

where $k$ is the magnitude of the propagation vector, and $\theta$ is the angle of incidence measured from the normal to the $xz$ plane (i.e. the plane containing the dipoles) – see Fig. 3. The dynamics of the system can then be modeled via equation (1). Substituting the ansatz for $x_n$ and $E_x$, namely:

$$\begin{cases} E_x = E_{x,0} \exp(ik_y y + ik_z nz_0 - i\omega t) \\ x_n = A_0 \exp(ik_y y + ik_z nz_0 - i\omega t) \end{cases} \quad (14)$$

the dynamical equation (1) becomes:

$$\left[\omega_0^2 - \omega^2 - i\omega\gamma + 2\sigma\cos(kz_0 \sin(\theta))\right] A_0 \exp(ik_z nz_0) = \frac{q}{m} E_{x,0} \exp(ik_z z) \quad (15)$$

Considering the field (12) in the points:

$$z = nz_0 \quad (16)$$

(15) becomes an equation for the amplitude $A_0$:

$$\left[\omega_0^2 - \omega^2 - i\omega\gamma + 2\sigma\cos(kz_0 \sin(\theta))\right] A_0 = \frac{q}{m} E_{x,0} \quad (17)$$

or:

$$A_0 = \frac{q}{m} \frac{E_{x,0}}{\left(\omega_0^2 - \omega^2 - i\omega\gamma + 2\sigma\cos(kz_0 \sin(\theta))\right)} \tag{18}$$

The polarization of the medium can thus be given by:

$$P_{x,0} = \eta q A_0 = \frac{\eta q^2}{m} \frac{E_{x,0}}{\left(\omega_0^2 - \omega^2 - i\omega\gamma + 2\sigma\cos(kz_0 \sin(\theta))\right)} \tag{19}$$

where $\eta$ is the concentration of the dipoles.

An effective susceptibility of the medium is defined via the equation:

$$P_{x,0}(\omega,k) = \eta q \chi(\omega,k) E_{x,0} \tag{20}$$

such that:

$$\chi(\omega,k) = \frac{q}{m} \frac{1}{\left(\omega_0^2 - \omega^2 - i\omega\gamma + 2\sigma\cos(kz_0 \sin(\theta))\right)} \tag{21}$$

The Helmholtz equation for the wave propagation can now be used. For the present case of dipoles $Q=0$, $M=0$, and using the plane wave ansatz, the Helmholtz equation is written as:

$$-\left(k_y^2+k_z^2\right)E_{x,0}+\frac{\omega^2}{c^2}\left(E_{x,0}+4\pi P_{x,0}(\omega,k)\right)=0 \tag{22}$$

Substituting the polar form for the wave vector components (13), the dispersion relation takes the final form:

$$k^2=\frac{\omega^2}{c^2}\left[1+4\pi\frac{q^2\eta}{m}\frac{1}{\left(\omega_0^2-\omega^2-i\omega\gamma+2\sigma\cos(kz_0\sin(\theta))\right)}\right] \tag{23}$$

This transcendental equation for $k$ has to be solved numerically. The results of the numerical solution of (23) are presented in Fig. 8.

### 3.2 Periodic chain of coupled quadrupoles

The following analysis is carried out using the definitions made in Section 2.2 (also see Fig. 4). To derive the dispersion relation for electromagnetic waves propagating in a medium with chains of coupled quadrupoles, the dynamics of system (5) has to be investigated under the influence of an electromagnetic plane wave. The dynamic equations can now be written as:

$$\begin{cases} \dfrac{\partial^2 x_n}{\partial t^2}+\gamma\dfrac{\partial x_n}{\partial t}+\omega_0^2 x_n+\sigma\left(x_{n+1}+x_{n-1}\right)+\sigma'\left(x'_{n+1}+x'_{n-1}\right)+\sigma_0 x'_n=\dfrac{qE_x}{m}\exp(ik_y y_1) \\ \dfrac{\partial^2 x'_n}{\partial t^2}+\gamma\dfrac{\partial x'_n}{\partial t}+\omega_0^2 x'_n+\sigma\left(x'_{n+1}+x'_{n-1}\right)+\sigma'\left(x_{n+1}+x_{n-1}\right)+\sigma_0 x_n=\dfrac{qE_x}{m}\exp(-ik_y y_1) \end{cases} \tag{24}$$

where the phase retardations are measured with respect to the center point **P** (see Fig. 4). Using the same ansatz:

$$\begin{cases} x'_n = A'_0 \exp(ik_y y + ik_z n z_0 - i\omega t) \\ x_n = A_0 \exp(ik_y y + ik_z n z_0 - i\omega t) \end{cases} \qquad (25)$$

and taking $z = n z_0$ as before, the dynamic equations can be written as:

$$\begin{cases} A_0 (R + 2\sigma \cos(k_z z_0)) + A'_0 (2\sigma' \cos(k_z z_0) + \sigma_0) = \dfrac{qE_x}{m} \exp(ik_y y_1) \\ A_0 (2\sigma' \cos(k_z z_0) + \sigma_0) + A'_0 (R + 2\sigma \cos(k_z z_0)) = \dfrac{qE_x}{m} \exp(-ik_y y_1) \end{cases} \qquad (26)$$

here $R = \omega_0^2 - \omega - i\gamma\omega$. Eventually, coordinates (25) are to be substituted into the relations for the polarization $P$, volume averaged quadrupole moment $Q$, and the magnetization $M$. Equations (26) have to be solved for $A_0$ and $A'_0$. This can be done by using the Cramer's method, which gives:

$$A_0 = \dfrac{\begin{vmatrix} \dfrac{qE_x}{m} \exp(ik_y y_1) & 2\sigma' \cos(k_z z_0) + \sigma_0 \\ \dfrac{qE_x}{m} \exp(-ik_y y_1) & R + 2\sigma \cos(k_z z_0) \end{vmatrix}}{\begin{vmatrix} R + 2\sigma \cos(k_z z_0) & 2\sigma' \cos(k_z z_0) + \sigma_0 \\ 2\sigma' \cos(k_z z_0) + \sigma_0 & R + 2\sigma \cos(k_z z_0) \end{vmatrix}} \qquad (27)$$

$$A_0 = \dfrac{qE_x}{m\Delta} \left[ (R + 2\sigma \cos(k_z z_0)) \exp(ik_y y_1) - (2\sigma' \cos(k_z z_0) + \sigma_0) \exp(-ik_y y_1) \right]$$

$$\Delta = (R + \sigma_0 + 2(\sigma + \sigma') \cos(k_z z_0))(R - \sigma_0 + 2(\sigma - \sigma') \cos(k_z z_0)) \qquad (28)$$

Similarly:

$$A_0' = \frac{\begin{vmatrix} \frac{qE_x}{m}\exp(ik_y y_1) & 2\sigma'\cos(k_z z_0)+\sigma_0 \\ \frac{qE_x}{m}\exp(-ik_y y_1) & R+2\sigma\cos(k_z z_0) \end{vmatrix}}{\begin{vmatrix} R+2\sigma\cos(k_z z_0) & 2\sigma'\cos(k_z z_0)+\sigma_0 \\ 2\sigma'\cos(k_z z_0)+\sigma_0 & R+2\sigma\cos(k_z z_0) \end{vmatrix}} \quad (29)$$

or:

$$A_0' = \frac{qE_x}{m\Delta}\left[(R+2\sigma\cos(k_z z_0))\exp(-ik_y y_1)-(2\sigma'\cos(k_z z_0)+\sigma_0)\exp(ik_y y_1)\right]$$
$$\Delta = (R+\sigma_0+2(\sigma+\sigma')\cos(k_z z_0))(R-\sigma_0+2(\sigma-\sigma')\cos(k_z z_0)) \quad (30)$$

For the symmetric mode:

$$A_0 + A_0' = \frac{qE_{x,0}}{m}\frac{2\cos(k_y y_1)}{(R+\sigma_0+2(\sigma+\sigma')\cos(k_z z_0))} \quad (31)$$

and for the asymmetric mode:

$$A_0 - A_0' = \frac{qE_{x,0}}{m}\frac{2i\sin(k_y y_1)}{(R-\sigma_0+2(\sigma-\sigma')\cos(k_z z_0))} \quad (32)$$

If the effective susceptibility is defined as:

$$\chi^{\pm}(\omega, k_z) = \frac{q}{m} \frac{1}{\left(R \pm \sigma_0 + 2(\sigma \pm \sigma')\cos(k_z z_0)\right)} \tag{33}$$

then following the framework of the multipole approach [2], the averaged multipole moments are:

$$\begin{cases} P_{x,0} = 4q\eta\chi^{+}(\omega, k_z)\cos(k_y y_1) E_{x,0} \\ Q_{x,0} = 2iq\eta y_1 \chi^{-}(\omega, k_z)\sin(k_y y_1) E_{x,0} \\ M_{z,0} = (-i\omega)2iq\eta y_1 \chi^{-}(\omega, k_z)\sin(k_y y_1) E_{x,0} \end{cases} \tag{34}$$

It can be shown that plugging the moments into the Helmholtz equation leads to the following relation:

$$k_y^2 + k_z^2 = \frac{\omega^2}{c^2}\left[1 + \chi^{+}(\omega, k_z)\cos(k_y y_1) + \chi^{-}(\omega, k_z) k_y y_1 \sin(k_y y_1)\right] \tag{35}$$

Substituting the polar form for the components (13), the dispersion relation takes the final form:

$$k^2 = \frac{\omega^2}{c^2}\left[1 + \chi^{+}(\omega, k\sin\theta)\cos(ky_1\cos\theta) + \chi^{-}(\omega, k\sin\theta) ky_1 \cos\theta \sin(ky_1\cos\theta)\right] \tag{36}$$

with:

$$\chi^{\pm}(\omega, k\sin\theta) = \frac{q}{m} \frac{1}{\left(R \pm \sigma_0 + 2(\sigma \pm \sigma')\cos(kz_0\sin\theta)\right)} \tag{37}$$

Equation (37) reveals the directional dependence of the material susceptibility. Dispersion relation (36)

represents the basic equation used to study the dispersion characteristics of the spatially dispersive MM with interaction between MA in lateral direction. To adhere to the methodology of the previous analysis [2] (i.e. in the study of MA without any form of interaction), the following approximation can be made:

$$\cos(k_y y_1) \approx 1 - (k_y y_1)^2, \quad \sin(k_y y_1) \approx k_y y_1 \tag{38}$$

and the propagation vector may be rewritten under this approximation as:

$$k^2 = \frac{\omega^2}{c^2} \left[ \frac{1 + A\chi^+(\omega, k\sin\theta)}{1 + \frac{\omega^2}{c^2} A y_1^2 \left( \frac{\chi^+(\omega, k\sin\theta)}{2} - \chi^-(\omega, k\sin\theta) \right)} \right] \tag{39}$$

This transcendental equation for $k$ has to be solved numerically. The results of the numerical solution of (39) are presented in Fig. 8.

### 4. Numerical solution of the dispersion relations

In what follows, the methods of analysis of the dispersion characteristics are jointly developed for the dipole and quadrupole systems. The dispersion characteristics are presented side by side to facilitate a comparative study of both systems, where (23) and (39) form the primary working equations for the above presented analysis. A modified version of the Regula Falsi method applicable to complex variables and commercial software MATLAB was used [8] to implement the numerical routine in order to solve (23) and (39).

#### 4.1 Verification of the computer code

The normalized forms of (23) and (39) were used. Specifically, (23) was reformulated as:

$$k^2 y_1^2 = \omega_n^2 \left( \frac{\omega_0^2 y_1^2}{c^2} \right) \left[ 1 + 4\pi \frac{q^2 \eta}{m\omega_0^2} \frac{1}{\left(1 - \omega_n^2 - i\omega_n/Q + 2C\cos(kz_0 \sin(\theta))\right)} \right] \quad (40)$$

Where $\omega_n = \omega/\omega_0$, $Q = \omega_0/\gamma$, $C = \sigma/\omega_0^2$, $z_n = z_0/(2y_1)$; $2y_1$, the cut wire spacing, was taken as 65 nm. The factor in the numerator $q^2\eta/(m\omega_0^2)$ was taken as $2,20 \times 10^{30}/4$ SI units. The resonant frequency of a single independent resonator was taken to be $\omega_0 = 1,39 \times 10^{15}$ rad/s, and the damping coefficient was taken as $\gamma = 9,42 \times 10^{13}$ rad/s. The values taken here were the same as in [3], where these parameters were obtained using the fitting with the rigorous computer simulations. To ensure a proper functioning of the computer code, the results of (40) at $\sin(\theta)=0$ were comparison with ones presented in [3]. Fig. 5 shows the wave vector and the effective material parameters obtained with $\sigma = 0,6 \times 10^{30}$ rad/s, $\gamma = 9,42 \times 10^{13}$ rad/s, and $z_n = 4,65$ corresponding to the values used in [3]. The large distance ($z_n = 4,65$) results in a very low value of coupling between MA. The obtained values match perfectly, confirming the correct functioning of the code.

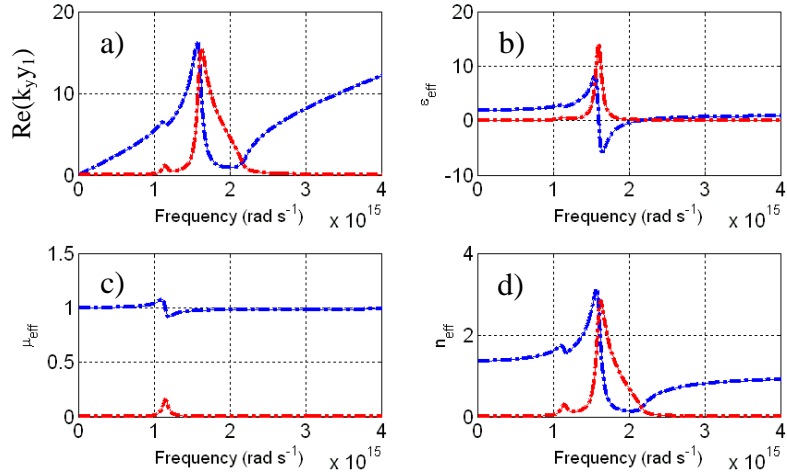

**Fig. 5: Verification of results obtained with code with the ones from [3]: (a) the propagation vector, (b) the effective permittivity, (c) the effective permeability, (d) the refractive index. The results from [3] are graphed with dotted lines. The results of the numerical code perfectly match the results and are indistinguishable from them.**

### 4.2 Results and discussions

With the correctness of the numerical procedure ensured, the focus is now shifted to the actual dispersion characteristics.

**a. Material eigen waves in the coupled dipole and quadrupole chains**

First of all, for the dipole chain the eigen value equation (4) in its normalized form is:

$$1 - \omega_n^2 - i\omega_n/Q + 2C\cos(k_z z_0) = 0 \qquad (41)$$

While for the quadrupole chain (11) after normalization becomes:

$$\begin{cases} 1 - \omega_n^2 - i\omega_n/Q + C + 2C(b+c)\cos(k_z z_0) = 0 \\ 1 - \omega_n^2 - i\omega_n/Q - C + 2C(b-c)\cos(k_z z_0) = 0 \end{cases} \qquad (42)$$

The propagation vector $k_z$ is treated as the independent variable, and (41), (42) are solved for the normalized frequency $\omega_n = \dfrac{\omega}{\omega_0}$ treating the normalized spatial period $u = k_z z_0$ as a parameter. The solutions of (41), (42) are obtained for different lateral coupling characterized by the normalized lateral distance $z_n = z_0/(2y_1)$ for two values of the damping constant, $\gamma = 0$ (see Fig. 6) and $\gamma = 9.42 \times 10^{13}$ rad/s (see Fig. 7). The real part gives the eigen frequencies that can sustain the spatial modes, while the imaginary part gives the time decay constant for the particular mode. The values of $u$ are limited to the range 0 to $2\pi$, as the solutions are periodic.

A widening of the band of eigen frequencies is observed as the spatial period decreases. Further, beyond a certain value for the periodicity, the medium starts to exhibit a stop band. The imaginary part of the solution remains independent of the propagation vector until the distance between the dipoles reaches the critical value and becomes $k$-dependent around the band stop.

For the quadrupoles, the solutions of the above equations are obtained for the symmetric $\omega_{n,symm}$ and asymmetric $\omega_{n,asymm}$ modes: the symmetric modes are in the higher band of eigen frequencies, while the asymmetric mode are for frequencies in the lower band. For both types of the modes, there is a band stop accompanied by a non-zero value for the imaginary part of the solution.

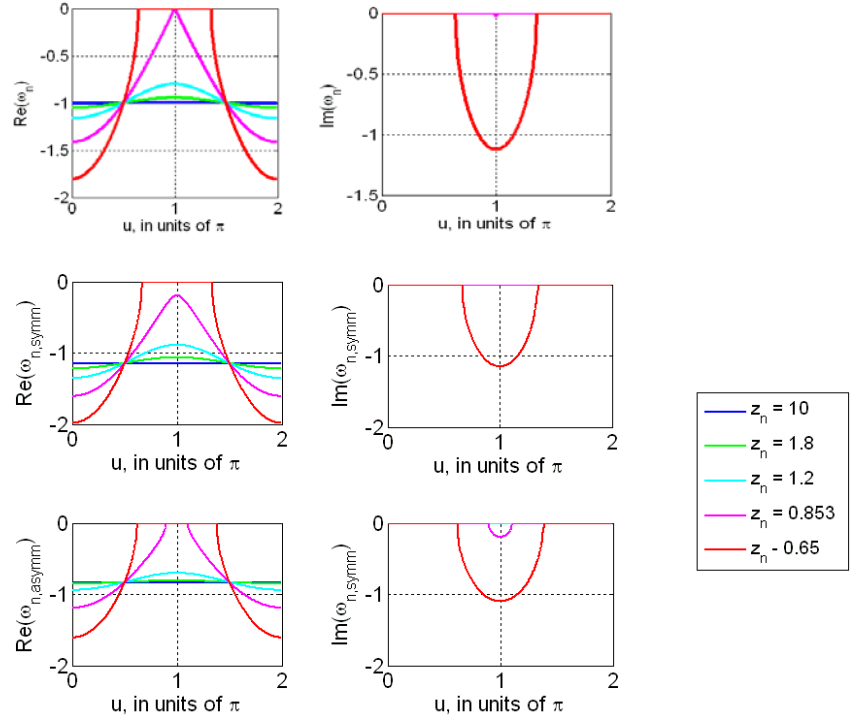

**Fig. 6:** Solution of the material eigen wave dispersion relation in media with chain of coupled dipoles and quadrupoles in the absence of material damping ($\gamma=0$). The normalized propagation vector $u = kz_0$ is treated as an independent variable, and the solutions are obtained in terms of the normalized eigen frequency $\omega_n$. The spatial period $z_n = z_0 / (2y_1)$ characterizes the coupling strength. The top row gives the real parts of the solutions, while the bottom row gives the imaginary part. (a), (d) - solution for dipole chain medium; (b), (e) - solution for symmetric mode in quadrupole chain medium; (c), (f) - solution for the asymmetric mode in quadrupole chain medium. The curves clearly indicate the onset of a stop band beyond a certain value of the coupling strength (lateral periodic spacing).

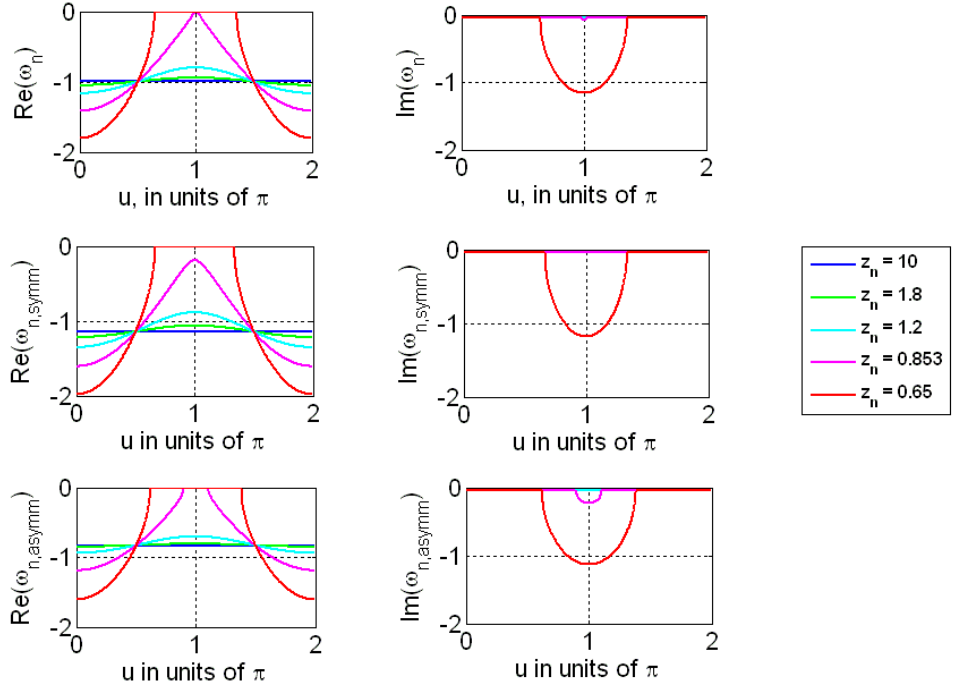

**Fig. 7:** Solution of the material eigenwave dispersion relation in media with chain of coupled dipoles and quadrupoles, in the presence of material damping ($\gamma \neq 0$). The normalized propagation vector $u = kz_0$ is treated as an independent variable, and the solutions are obtained in terms of the normalized eigen frequency $\omega_n$. The spatial period $z_n = z_0/(2y_1)$ characterizes the coupling strength. The top row gives the real parts of the solutions, while the bottom row gives the imaginary part. (a), (d) - solution for dipole chain medium; (b), (e) - solution for symmetric mode in quadrupole chain medium; (c), (f) - solution for the asymmetric mode in quadrupole chain medium. There are no significant changes in the real part of the solutions, but the imaginary parts are now slightly shifted downwards owing to the presence of the material damping.

### b. Dispersion relation for electromagnetic waves in coupled-dipole and quadrupole chain medium

The dispersion relations for the electromagnetic wave propagating in a media with the coupled dipoles and quadrupoles (23) and (39) in their normalized forms are:

$$u^2 = k^2 y_1^2 = \omega_n^2 \left( \frac{\omega^2 y_1^2}{c^2} \right) \left[ 1 + 4\pi \frac{q^2 \eta}{m\omega_0^2} \frac{1}{\left(1 - \omega_n^2 - i\omega_n/Q + 2C\cos(kz_0 \sin(\theta))\right)} \right] \quad (43)$$

$$u^2 = k^2 y_1^2 = \frac{\omega^2 y_1^2}{c^2} \left[ \frac{1 + A\chi^+(\omega, k\sin\theta)}{1 + \frac{\omega^2}{c^2} Ay_1^2 \left( \frac{\chi^+(\omega, k\sin\theta)}{2} - \chi^-(\omega, k\sin\theta) \right)} \right] \quad (44)$$

where:

$$\chi^\pm(\omega, k\sin\theta) = \frac{q}{m\omega_0^2} \frac{1}{\left(1 - \omega_n^2 - i\omega_n/Q + \pm C + 2C(b\pm c)\cos(kz_0 \sin(\theta))\right)} \quad (45)$$

with: $z_n = z_0/(2y_1)$, $\omega_n = \omega/\omega_0$, $b = 1/z_n^3$, $c = 1/(1+z_n^2)^{3/2}$, $Q = \omega_0/\gamma$. Solution of dispersion relations (43), (44) were found numerically for two different periodicities $z_n = 1,8$ and $z_n = 1,2$. The reference spacing was taken to be $2y_1 = 65nm$. For each of these two periodicities, the dispersion characteristics were calculated for two values of the damping coefficient $\gamma = 9,42 \times 10^{13}$ rad/s and $\gamma = 5 \times 10^{13}$ rad/s. This resulted in overall four sets of data. The results for both dipoles and quadrupoles are given together in Fig. 8, which shows a set of dispersion curves obtained for different angles of incidence. The difference between the dispersion curves for dipoles and quadrupoles is the presence of another set of spatial modes corresponding to quadrupoles, at a frequency lower than the dipole resonance frequency (see for e.g., Fig. 8 (c), (d)). However, the response from the quadrupoles is much smaller. Also, the response at these lower frequencies completely disappears for incidence at $\theta = \pi/2$ - i.e. when light is propagating along the $z$ direction. No quadrupolar moments are excited at this angle as both the cut wires were excited by the same field. The width of the band gap is different in the two systems, which is caused by the fact that in quadrupoles each meta-atom comprises of two cut-wires, giving rise to the enhanced absorption width. It is further observed that the angle of incidence has a clear and pronounced effect on the position of the symmetric response. Specifically, the resonance peak shifts towards the blue part of the spectrum as the angle of incidence increases from $0$ to $\pi/2$, while it is red shifted for the imaginary part (see for e.g., Fig. 8 (g), (h)).

The curves of Figs. 9, 10, 11 show the variation of the effective material parameters (permittivity, permeability and the refractive index) with the incident angle. The trends as observed for the propagation vector are replicated here as well. The permeability goes to zero at all wavelengths for an incident angle of $\pi/2$ (Fig. 10). Fig. 11 shows the variation of the effective material parameters with the spatial periodicity. In the curves, the frequency of maximum response, and the corresponding value are plotted as a function of the normalized spatial period $z_n$. The angle of incidence is treated as a parameter. Again, the results of this analysis at a periodicity of $z_n = 4, 6$ for $\theta = 0$ matched perfectly with the results obtained in [3].

In the first row (Fig. 12(a), (b), (c)), the resonant frequencies of the effective parameters are plotted as a function of the normalized spatial period $z_n$. The resonant frequencies for the respective effective parameters asymptotically approach to the values obtained in [3]. This asymptotic tendency indicates that as the spatial period becomes larger, the near field interactions between the MA become weaker, and hence the quadrupoles become decoupled as assumed in [3]. It is interesting to note that the position of the antisymmetric resonance depends on the spatial period. In the second row (Fig. 12(d), (e), (f)), the values of the imaginary parts of the respective parameters at resonance are plotted as a function of the spatial period. At $z_n = 5$, the peak values match those as obtained in [3]. These correspondences indicate the correctness of the presented results.

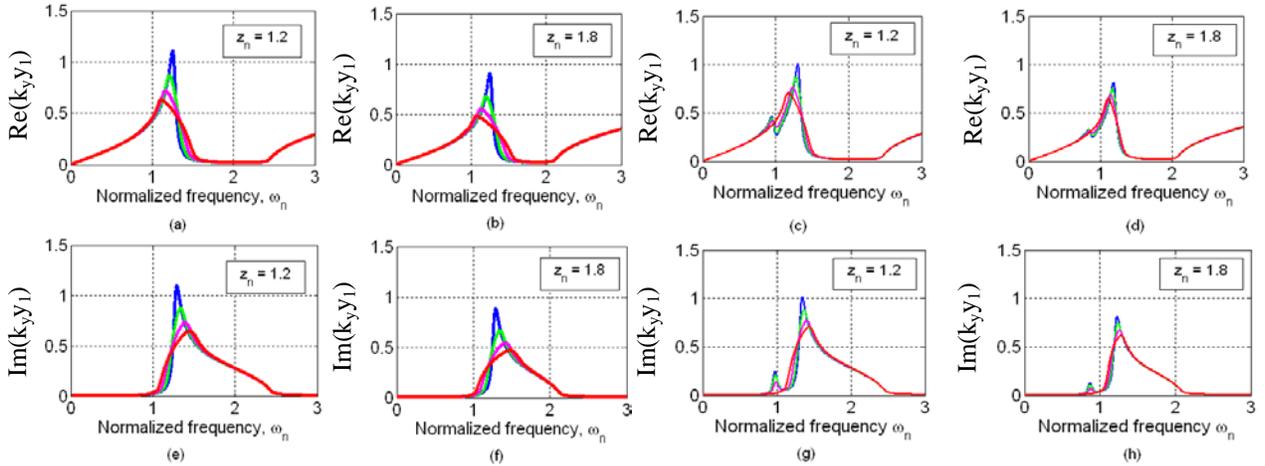

**Fig. 8: Electromagnetic dispersion curves for a system consisting of one dimensional chain of coupled dipoles and quadrupoles for two spatial periods. (a), (b), (e) and (f) depict the dispersion relations for the dipole system, while (c), (d), (g), and (h) depict the dispersion relations for the quadrupole system. The first row depicts the real part of the normalized propagation vector $k_y y_1$, while the bottom row depicts the imaginary part. The values were obtained with**

the incident angle as parameter (blue – 0, green – $\pi/8$, cyan – $\pi/4$, red – $\pi/2$). Note the disappearance of the resonance associated with the quadrupole and magnetic dipole moment at the incident angle of $\pi/2$.

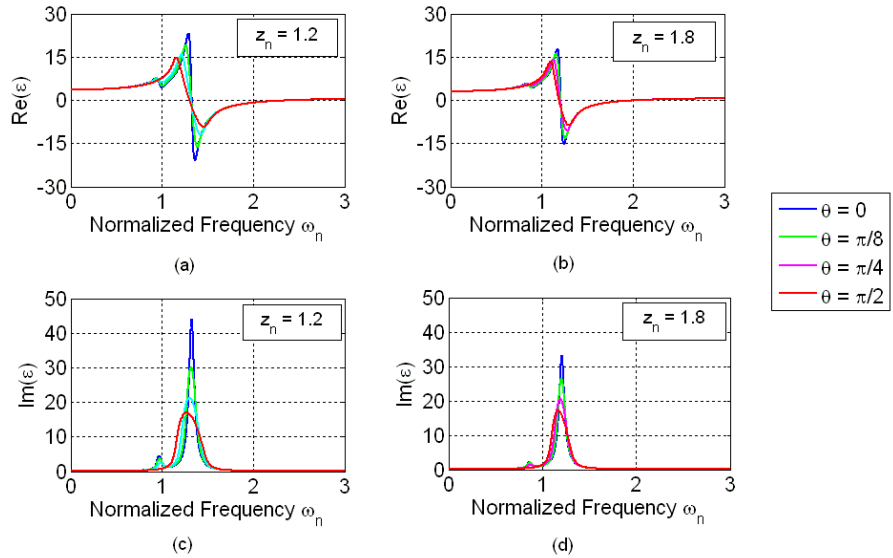

**Fig. 9: Effective permittivity for quadrupoles for different angles of incidence - the curves are obtained for the two periodicities: $z_n = 1,2$ and $z_n = 1,8$. The first row depicts curves for the real value of the effective permittivity, while the second row gives the imaginary parts of the effective permittivity. (a) and (c) were obtained for $z_n = 1,2$, while (b) and (d) were obtained for $z_n = 1,8$.**

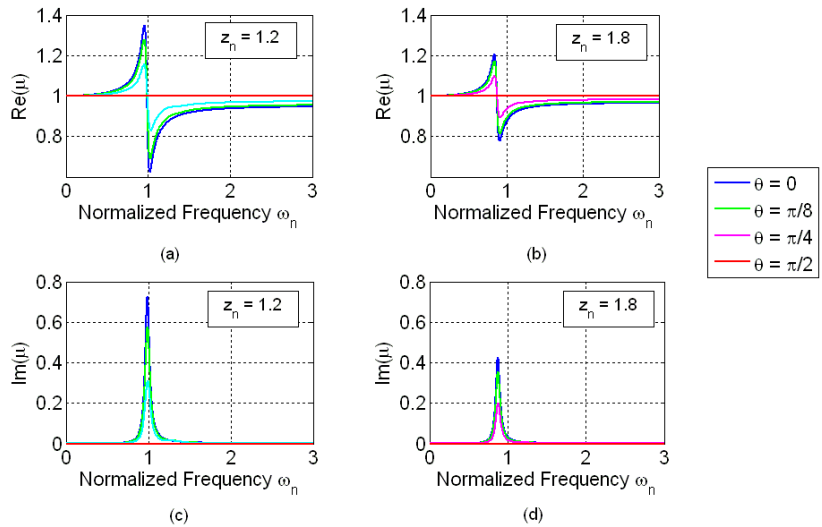

**Fig. 10: Effective permeability for quadrupoles for different angles of incidence - the curves are obtained for the two periodicities: $z_n = 1,2$ and $z_n = 1,8$. The first row depicts curves for the real value of the effective permeability, while**

the second row gives the imaginary parts of the effective permeability. (a) and (c) were obtained for $z_n = 1,2$, while (b) and (d) where obtained for $z_n = 1,8$.

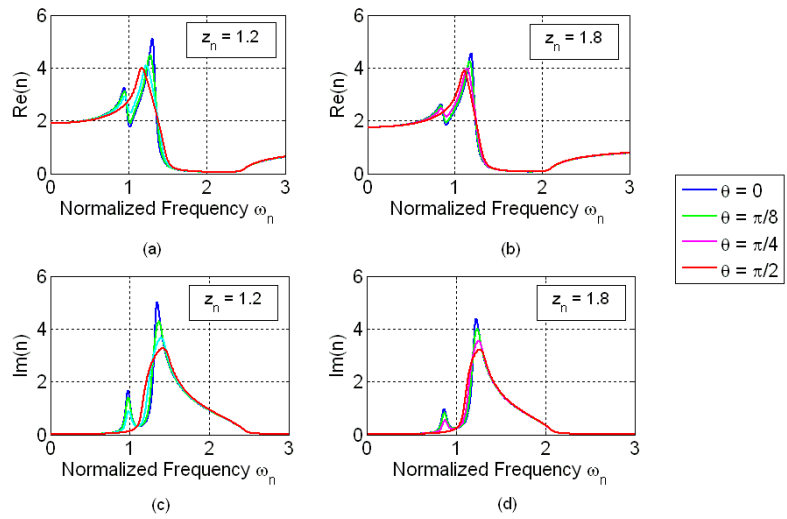

**Fig. 11:** Effective refractive index for quadrupoles for different angles of incidence - the curves are obtained for the two periodicities: $z_n = 1,2$ and $z_n = 1,8$. The first row depicts curves for the real value of the effective index, while the second row gives the imaginary parts of the effective index. (a) and (c) were obtained for $z_n = 1,2$, while (b) and (d) were obtained for $z_n = 1,8$.

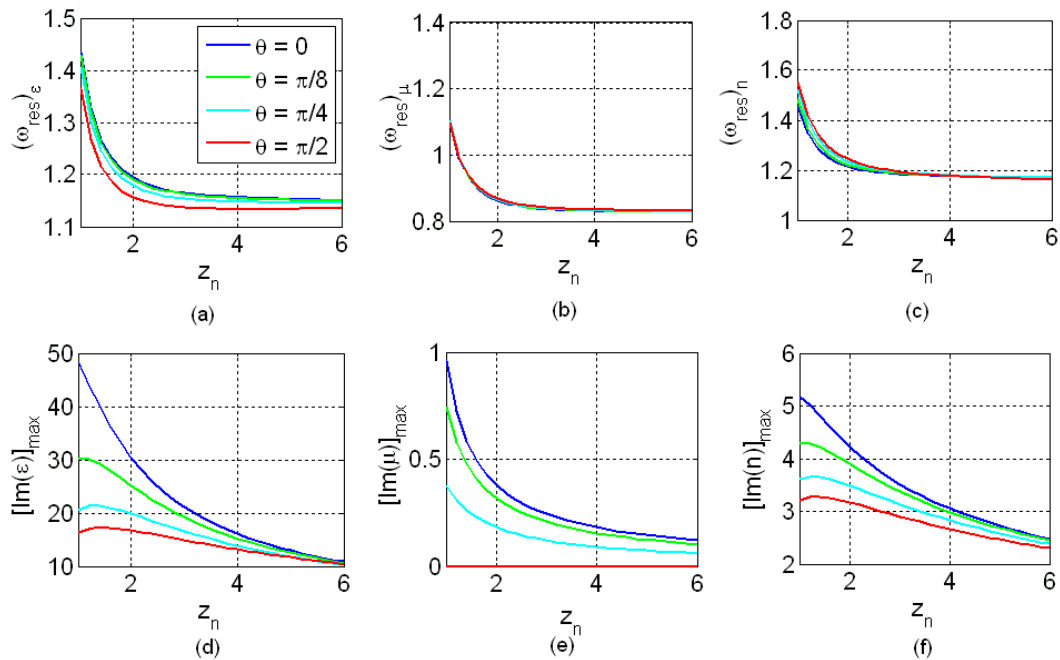

**Fig. 12:** Material parameter curves as a function of period of spacing $z_n$ - the top row are plots of resonant frequency as a function of the spatial periodicity, while the bottom row are the plots of the values of the imaginary

**part of effective material parameters at the resonant frequencies as a function of $z_n$ : (a), (d) were obtained for the imaginary part of the effective permittivity, (b), (e) for the imaginary part of the effective permeability, and (c), (f) for the imaginary part of the refractive index.**

### c. Analysis of the spatial spectrum available for the potential optical resolution enhancement

The available spatial spectrum, given by a dispersion relation for particular media, determines maximum resolution, which could be achieved with this media. Explanation of this statement can be found in any university textbook; qualitative understanding is in fact, that the richer the available spatial spectrum the higher resolution could be provided due to the evident requirement of the Fourie transformation. Hence, roughly speaking, the higher values of the available *k* vectors are obtained from (23) and (39), the higher resolution could be achieved with the media possessing the dispersion relation (23) or (39). The solutions of these equations are given in Fig. 8 for both dipoles and quadrupoles cases. Comparing Figs. 8 a) and b) for different coupling strengths reveals that for the higher coupling (Fig. 8 a) the maximum available value of the real part of the *k* vector is higher than for the case of lower coupling (Fig. 8 b). The same relation – higher available *k* vectors values for the higher coupling – takes place for the quadrupoles as well, see Figs. 8 c) and d). Nevertheless, this enhancement of the real parts of the wave vectors does not allow us to conclude that this effect could lead to the respective resolution enhancement. The problem is in that the increase of the maximum available real parts of the wave vectors is accompanied by approximately the same percentage of increase of the losses, i.e. increase of the imaginary parts of the wave vectors, which lead to the suppression of the effect of higher resolution [19].

### 5. Conclusion

The presented here analytical treatment focused on the extension of the multipole model of MM to the case of MM with interacting MA, where there is a significant coupling between neighboring MA in lateral direction. The coupling gives rise to a response that is non-local - that is, the response is mediated by not only the response of a single meta-atom, but also their coupling with the immediate neighborhood. In such a case, spatial modes can be sustained in the MM; excitation the medium at exactly these eigen modes promises enhanced interaction with the MM. The above analysis confirms the existence of such spatial modes, and throws light on their characteristics. Two resonances occur - corresponding to the antisymmetric and symmetric modes of oscillation of the charges in the MA. The positions of the resonances can be controlled by changing the spatial periodicity of placement of the

MA. In the limit of large interspacing periods, the values of resonant frequencies obtained here correspond to the original analysis [3] for non-interacting MA, indicating the validity of the analysis. The dispersion relation changes caused by the lateral interaction between MAs are not expected to be used for the resolution enhancement properties due to the fact, that the real part wave vector enhancement is accompanied by the increase of losses, which suppress the effect. Further investigation of the interaction between the MAs in the longitudinal direction has to be performed in order to obtain full analysis of the influence of the interaction between the MAs on the dispersion properties of the MMs.

## References

[1] A. Chipouline, C. Simovski, S. Tretyakov, "Basics of averaging of the Maxwell equations", ArXiv 1205.2853, http://arxiv.org/abs/1205.2853.

[2] A. Chipouline, J. Petschulat, A. Tuennermann, T. Pertsch, C. Menzel, C. Rockstuhl, F. Lederer, „Multipole approach in electrodynamics of MM", Appl. Phys. A, 103, 899-904 (2011).

[3] J. Petschulat, C. Menzel, A. Chipouline, C. Rockstuhl, A. Tünnermann, F. Lederer, and T. Pertsch, "Multipole approach to MM," Phys. Rev. A 78, 043811 (2008).

[4] V. Shalaev, W. Cai, U. Chettiar, H-K. Yuan, A. Sarychev, V. Drachev, A. Kildishev, "Negative index of refraction in optical MM", Optics Letters, 30, 3356, 2005.

[5] P. Mazur, B.R.A. Nijboer, On the statistical mechanics of matter in an electromagnetic field. I, *Physica XIX*, pp. 971-986 (1953).

[6] A. Nikolaenko, N. Papasimakis, A. Chipouline, F. De Angelis, E. Di Fabrizio, N. Zheludev, "THz bandwidth optical switching with carbon nanotube metamaterial", OPTICS EXPRESS, 20, 6, 6068, 2012.

[7] A. Chipouline, V. A. Fedotov and A. E. Nikolaenko, "Analytical Model for MM with Quantum Ingredients", ArXiv 1104.0110 (2011)

[8] E. Clayton and G. H. Derrick, "A numerical solution of wave equations for real or complex Eigenvalues", Aust. J. Phys., vol. 30, p. 15 (1977).

[9] C. Simovski, „Material parameters of MM (a Review)", Optics and Spectroscopy, 107, 726, 2009.

[10] C. Simovski, "On electromagnetic characterization and homogenization of nanostructured MM", J. Opt. 13, 013001, 2011.

[11] A. Sarychev, V. Shalaev, "Electrodynamics of MM", World Sci., Singapore, 2007.

[12] D. Smith, D. Shurig, Phys. Rev. Lett., 90, 077405, 2003.

[13] P. Belov, C. Simovski, Phys. Rev. E 72, 026615, 2005.

[14] E. Tatartschuk, A. Radkovskaya, E. Shamonina, and L. Solymar, "Generalized Brillouin diagrams for evanescent waves in metamaterials with interelement coupling", Phys. Rev. B **81**, 115110, 2010.

[15] A. Radkovskaya, E. Tatartschuk, O. Sydoruk, E. Shamonina, C. J. Stevens, D. J. Edwards, and L. Solymar, "Surface waves at an interface of two metamaterial structures with interelement coupling", Phys. Rev. B **82**, 045430, 2010.

[16] A. Radkovskaya, O. Sydoruk, E. Tatartschuk, N. Gneiding, C. J. Stevens, D. J. Edwards, and E. Shamonina, "Dimer and polymer metamaterials with alternating electric and magnetic coupling", Phys. Rev. B **84**, 125121, 2011.

[17] E. Shamonina, "Magnetoinductive polaritons: Hybrid modes of metamaterials with interelement coupling", Phys. Rev. B **85**, 155146, 2012.

[18] Z. Jacob, L. Alekseev and E. Narimanov, "Optical Hyperlens: Far-field imaging beyond the diffraction limit", Optics Express, 14, 2006.

[19] D. Smith, D. Schurig, M. Rosenbluth, S. Schultz, S. Ramakrishna, and J. Pendry, "Limitation on subdiffraction imaging with a negative refractive index slab", APL, 82, 1506, 2003.